# EEG-connectivity: A fundamental guide and checklist for optimal study design and evaluation

Aleksandra Miljevic[1]*, Neil W. Bailey[1], Fidel Vila-Rodriguez[2], Sally E. Herring[1], Paul B. Fitzgerald[1]

[1]Epworth Centre for Innovation in Mental Health, Department of Psychiatry, Central Clinical School, Monash University, Epworth HealthCare, 888 Toorak Rd, Camberwell, Victoria 3124, Australia.

[2]Non-Invasive Neurostimulation Therapies Laboratory, Dept. Psychiatry, The University of British Columbia, Vancouver, BC, Canada.

*Corresponding author

Abstract

Brain connectivity can be estimated through a wide number of analyses applied to electroencephalographic (EEG) data. However, substantial heterogeneity in the implementation of connectivity methods exist. Heterogeneity in conceptualization of connectivity measures, data collection, or data pre-processing may be associated with variability in robustness of measurement. While it is difficult to compare the results of studies using different EEG connectivity measures, standardization of processing and reporting may facilitate the task. We discuss how factors such as referencing, epoch length and number, controls for volume conduction, artefact removal, and statistical control of multiple comparisons influence the EEG connectivity estimate for connectivity measures, and what can be done to control for potential confounds associated with these factors. Based on the results reported in previous literature, this article presents recommendations and a novel checklist developed for quality assessment of EEG connectivity studies. This checklist and its recommendations are made in an effort to draw attention to factors that may influence connectivity estimates and factors that need to be improved in future research. Standardization of procedures and reporting in EEG connectivity may lead to EEG connectivity studies to be made more synthesisable and comparable despite variations in the methodology underlying connectivity estimates.

**Highlights**

- Research findings, background information, and recommendations from the existing EEG literature are summarised and compiled to propose a novel checklist to evaluate EEG connectivity analyses.
- The checklist can be used in both the developmental stages of a study (i.e., when choosing which methods to use) as well as when assessing published studies (i.e., when assessing studies for a meta-analysis).
- The checklist is made in an effort to draw attention to the existing methodological gaps and inconsistencies in EEG research, so that future connectivity estimates and the factors influencing them may be standardised and improved.

The human brain operates as a network of functionally interconnected regions; the ‘connectivity’ or synchronised activity of which is believed to underpin behaviour, cognition, and mood states (Anderson et al., 2016). Therefore, these networks and the connections within and between them are important brain features to understand.

Electroencephalography (EEG) is a low-cost and low-burden tool that can measure the electrical activity of the brain with high temporal resolution (Jackson & Bolger, 2014; Michel, 2009). EEG activity oscillates, with voltages shifting from negative to positive and back again, multiple times per second. As a result of decades of research, information recorded by EEG is believed to be transmitted through oscillatory synchronisation of the brains neurons (Fries, 2005). These oscillations can be interpreted either in: the time-domain, which allows measurements of absolute voltage changes across time (i.e., as event-related potentials; Cohen, 2014); in the frequency-domain, which measures the amplitude of oscillations within specific frequency bands; or the time-frequency domain, where changes in the phase and patterns of different frequencies are assessed across time (Cohen, 2014). Typically, EEG signals are distinguished into five frequency bands: delta, theta, alpha, beta and gamma, within which patterns of frequency power (the amplitude of the voltage fluctuations within a specific oscillatory frequency) and peak frequency are typically assessed (Michel, 2009).

‘Connectivity’ in EEG (hereafter referred to as ‘EEG-connectivity’) involves the consideration of the relationship between two or more EEG signals (Cohen, 2014; Michel, 2009). In this way, EEG is a tool through which connectivity within the brain can be assessed either in time, frequency, or time-frequency domains.

## 1 A brief introduction to EEG-connectivity

Assessing connectivity with EEG is possible, informative, and growing in popularity. As mentioned, the approach most used in research assessing connectivity through EEG is the identification of statistically significant synchronisation (initially through correlational approaches, and more recently, through complex analyses) between the signals obtained from two or more EEG electrodes (Cohen, 2014). There are several techniques or strategies that can be used to achieve connectivity analyses. We start discussing key concepts and definitions. For the reader's own reference, additional definitions of key terms used throughout the manuscript are included in Supplementary Material A: Key Term Definitions.

### 1.1 Scalp versus source -space connectivity

Signal, sensor, or scalp -space connectivity is where EEG-connectivity is estimated based on activity potentials obtained from the scalp (i.e., through the electrodes, without attempting to determine where the underlying activity is generated from). However, it is possible to transform this scalp-space EEG data to estimate the location and distribution of the signals, to identify 'sources' responsible for the observed activity (Jatoi et al., 2014). This is commonly referred to as 'source-localisation' or 'source-space' activity, which can be achieved either through network structures and modelling, or with model-free techniques [see Grech et al. (2008), Jatoi et al. (2014) Mahjoory et al. (2017) for a review of methods].

Source-localization approaches define sets of weights per electrode, and the weighted sum of all electrodes is an estimate of activity originating from some physical location in the brain (Michel et al. 2004). This weighting can be achieved through two approaches: 1) the forward solution, where an estimate of a topographical map that would result from the activity of a recorded dipole in a specific region of the brain with a specific orientation, is made; and 2) an estimated solution to the inverse problem, where the most likely dipole locations, orientations and magnitudes that could have produce the observed results topography are estimated (Cohen et al., 2009; Dominguez et al., 2017).

Typically, the inverse problem is utilised most frequently in EEG source-localisation. This technique makes assumptions about the electrophysiological and neuroanatomical constraints of the brain's grey matter, which have been informed by Magnetic Resonance Imaging (MRI) research. From these assumptions, the electrical activity generated by the cortex can be modelled as a collection of voxels (or volume elements), for which we know the orientation and strength, of connections between neighbouring neural populations, based on the MRI information constraints around grey matter.

While a useful methodology, there is no single solution to the inverse problem, source-localisation estimates are complex and pose difficulties including that calculation of source activity relies on assumptions that are not able to be tested directly in the data. Thus, researchers cannot infer with certainty the source locations and number of generators of the signal based on the recorded EEG data.

### 1.2 Bivariate versus multivariate connectivity

Another important distinction within the assessment of EEG-connectivity is that the analyses can be performed as: 1) bivariate analyses, either in *pairs* (electrode-to-electrode or source-to-source), or *globally* (where the bivariate connectivity values are averaged to produce one overall value); or 2) multivariate analyses (electrode-to-electrode-to-electrode, source-to-source-to-source, or more), using approaches like graph theory, a mathematical framework for characterizing interconnected networks (Cohen, 2014). Most studies assess bivariate connections. While both bivariate and multivariate methods have their limitations and benefits, multivariate analyses of brain connectivity are still underdeveloped – they are vague in their depiction and assessment of connections and their strengths, which can result in difficulty with interpreting the significant connections within a network (Cohen, 2014).

### 1.3 Over-time versus across-trial connectivity

The focus of this guide is measures of connectivity between different electrodes and by extensions, the underlying brain regions. However, it is worth noting that connectivity is often referred to as ‘synchronisation’, a term which is commonly used to refer not just to connectivity between electrodes or brain regions, but also synchronisation between the phase of oscillations and the presentation of stimuli over multiple presentation of that stimuli. This has been referred to as ‘inter-trial coherence’, or across-trial connectivity. This can create further confusion, as these terms are also used to refer to connectivity between electrode / brain region analyses.

It should be noted that most connectivity measures can be adjusted to be assessed as either over-time or over-trial, rather than between electrodes or brain regions. Across-trail measures seem to be used most when connectivity is assessed during cognitive processing.

Lastly, while it is helpful to be aware that the distinction between these two analyses exists, across-trail measures will not be discussed further in this guide, which is focused on connectivity between electrodes / brain regions across time.

### 1.4 Power versus phase -based connectivity

In general, power-based connectivity measures first convert the data into the time-frequency domain, then assess whether there is a relationship in changes to power of a specific oscillation between two different electrodes or brain regions. In contrast, phase-based connectivity measures assess whether the phase angle of voltage shifts is related between two electrodes or brain regions. While not enough research exists to suggest when power or phase-based connectivity measures are optimal, phase-based measures are suggested to be less sensitive to spurious interaction in the EEG; meaning interactions between electrodes that are driven by an artifact of the EEG recording or analysis, rather than underlying brain activity. This is because they ignore the zero phase or instantaneous interactions thought to be the result of volume conduction (VC) and do not rely on the amplitude of the signal (van Diessen et al., 2015; Muthukumaraswamy & Singh, 2011). Overall, the two measures also tend to reveal different result patterns, both due to their mathematical perspectives and the fact that they are thought to reflect different neurophysiological processes.

Phase-based connectivity is thought to reveal the timing of activity within neural populations, whereas power-based connectivity is thought to reveal the number of neurons or the spatial extent of the neural populations (Cohen, 2014). Generally, phase-based connectivity measures are more commonly used in the literature and there is suggestion that phased-based measures are useful for hypotheses concerning instantaneous connectivity. However, power-based measures are more robust to temporal offset and jitter. Both power and phase -based connectivity can be assessed either over trial or time, using bivariate or multivariate analyses, and in the scalp or source -space. As such, the choice of which measure depends on the hypotheses and researchers aims proposed.

### 1.5 Effective versus functional connectivity

Connectivity can be estimated either as: *effective* connectivity, which is unidirectional, and attempts to determine the casual flow of information from one point to another (i.e., if activity in one brain region precedes activity in another region in time; Friston, 2011); or *functional* connectivity which is bi-directional and cannot determine causation (i.e., are two brain regions sharing common activity, suggesting they are connected; Friston, 2011). Research on both exists, with functional connectivity seeming to be used more often, possibly because it has been found to be a more statistically robust measure (Cohen, 2014).

### 1.6 Resting-state versus task-related connectivity

Connectivity can be assessed either while participants are at rest (not engaging in any goal- or task- orientated cognitions) or during the performance of a specific cognitive task. If connectivity is assessed in the resting-state, there are two conditions under which activity can be recorded: eyes-closed or eyes-open. In recent years, eyes-open conditions have been used less commonly, as several papers have emerged indicating that eyes-closed resting-state conditions provide more sensitivity to detect effects of interest in brain activity assessments (van Diessen et al., 2015).

Lastly, while studies assessing connectivity during cognitive task performance (also referred to as *task-related connectivity*) offer the opportunity to compare connectivity that is related to specific cognitive functions, there are currently fewer task-related than resting-state connectivity studies. Further, as will be discussed later in the article, task-related connectivity measures present some additional and unique challenges to the estimation of connectivity via EEG.

## 2 Assessing connectivity with EEG

While some connectivity methods have been found to be clearly inferior to others (increasing the probability of false positive or false negative results), there is no single method or best technique with which EEG-connectivity can be quantified and assessed. Wang et al. (2014) identified 42 methods for calculating EEG-connectivity (for further reviews on connectivity measures, see Sakkalis, 2011; Bakhshayesh et al., 2019), and more measures for assessing connectivity continue to be developed (for examples, see Mamashli et al., 2019; Wu et al., 2017). Each method assesses connectivity in its own unique way, often only with minor differences between a newly developed method and previously existing methods, and sometimes with completely different underlying theoretical and practical adaption. In some cases, this variability makes it difficult to reliably compare difference between studies using varying estimates. This is further complicated by the lack of standardisation in the EEG recording and processing steps that precede the connectivity computations. A growing body of research literature has demonstrated that non-optimal choices in EEG data analysis can led to biases and increased rates of false positives in connectivity measurements (Bastos & Schoffelen, 2016; Bakhshayesh et al., 2019). Thus, to be able to understand the differences between varying estimates of EEG-connectivity and the underlying brain connectivity they claim to assess, we first need to understand how the preceding steps affect the EEG-connectivity estimate and aim to standardise them.

Based on the existing literature, the following specific steps/variables have been identified as necessary and most appropriate to address to meet this aim: 1) how to control for noise and artifact removal; 2) how/if VC will be controlled for; 3) referencing; and 4) epoching parameters (including length and number of epochs). In addition to these four steps, network theory use and assessment, assumptions around statistical testing, and a discussion of the importance of sample size are also important to consider and highlight. We believe that the methods and statistics used to estimate connectivity need to be presented with a clear rationale (explaining how and why the chosen measure best suits the data), assessments of sample size should be stated, and explanations provided as to how multiple comparisons are controlled for. These points are necessary to address in connectivity research if useful comparisons with other studies are to be made, and our understanding of connectivity advanced (Chella et al., 2016; Cohen, 2014; 2015; van Diessen et al., 2015; Friston, 2013).

Currently, published studies are not consistently controlling for these variables or reporting how they have been controlled for. Therefore, in this paper we aimed to outline the critical issues, and present a checklist identifying the key components. The goal of this guide and the checklist is to highlight specific and key methodological guidelines for consideration prior to commencing an EEG-connectivity study or analysis, when reporting results, and when assessing the quality and interpreting the results of a published EEG-connectivity study. Each methodology is discussed in the following sections with brief summaries at the end. This paper will primarily focus on the 'resting-state' EEG measures, as these have been most researched. However, much of the information is applicable to task-related EEG and where feasible, differences between the two are discussed.

### 2.1 The effects of artifacts, and processing steps to remove artifacts

EEG recordings are a combination of useful brain-related information and 'noise' information (produced by non-brain related 'artifacts' e.g., electrical potentials produced by eye blinks, head muscle activation, electrical interference, and electrode displacement). When these noise artifacts are recorded by two neighbouring electrodes concurrently and analysed, the electrodes may produce a high estimate of connectivity. However, this is an incorrect identification of connectivity, as the estimate of connectivity is based on the noise rather than the underlying brain connections (see Bastos & Schoffelen, 2016, for specific examples). Thus, it is extremely important to base EEG-connectivity estimates on data that contains as little noise as possible for more accurate results, irrespective of whether scalp- or source- level analyses are applied, as both can be affected by artifacts.

Perhaps most usefully the studies should aim to reduce electrical impedances, minimisation of environmental electrical noise, and briefing participants on the negative impact of muscle activity and movement in the EEG recording session. Then, the remaining artifact can be removed in the EEG data pre-processing and transformation stages (Keil et al., 2014). There are three broad categories to the post recording noise/artifact rejection techniques: 1) manual rejection; 2) automated rejection; and 3) semi-automated rejection (the specific details of each of the techniques is beyond the scope of this article, for more see Gabard-Durnam et al., 2018).

Further, there are two approaches to addressing noise/artifact that are used across these categories: 1) rejecting segments of EEG data or specific EEG electrodes that contain artifacts; and 2) using mathematical methods to reduce or subtract the influence of the artifact from the data, while keeping the epochs and electrodes [common techniques include Independent Component Analysis (ICA, see Delorme et al., 2007; Pester & Ligges, 2018), Artifact Subspace Reconstruction (see Chan et al. (2019), or Weiner filters (see Somers et al., 2018)]. However, artifact rejection methods are imperfect, and some artifact related activity is likely to remain – the aim is simply to minimize the influence of this artifact on the connectivity results.

At its most basic, the ICA technique decomposes the EEG signal into statistically separate independent components, which can be categorized as neural and non-neural components (e.g., eye blinks, muscle movements, and heart beats), these non-neural components can then be subtracted, and the overall EEG trace can be reconstructed to the electrode space, 'corrected' for non-neural components to reflect this (Issa & Juhasz, 2019). However, ICA-corrected EEGs have been suggested to produce distortions in amplitude and phase which lead to spurious hyper-connectivity in studies assessing coherence-based estimates of connectivity across all frequency bands, and therefore false positive conclusions (Castellanos & Makarov, 2006). In contrast, wavelet enhanced ICA (wICA; aims to reduce artifact activity within an independent component, instead of subtracting the entire component; Issa & Juhasz, 2019), has been found to produce less distortions in amplitude and phase, resulting in better coherence estimation (Castellanos & Makarov, 2006).

Therefore, an artifact cleaning technique that does not subtract components rather, reduces them such as the wICA, is recommended when cleaning EEG data for connectivity analysis (it is worth noting that Castellanos and Makarov (2006) applied wICA to all components regardless of whether they were brain or artifact related, but wICA can be applied only on components identified as artifacts, which could preserves the characteristics of neural activity in the processed data more effectively). Multiple Wiener filtering can also be used to reduce non-neural activity. These filters aim to filter out rather than subtract noise from a signal (Huang et al., 2008), but are yet to be tested in EEG-connectivity studies, thus require further exploration and methodological research as to their effects on connectivity estimates. Whichever artifact reduction method is used, the steps need to be performed according to state of the art recommendations, and clearly reported. This is a necessary step to confirm that connectivity analyses are not adversely affected by artifacts, and to enable replication of study connectivity results.

While there are several artifact rejection techniques in use, depending on the EEG data processing used, the cleaning effectiveness may vary, with more signal or noise being retained (see Gabard-Durnam et al., 2018 for specifics). These techniques have been investigated thoroughly in the general EEG literature but there is little to no evidence on the impact they have in EEG-connectivity. Manual rejection is currently the most common method, however a problematic one for the research field, where standardisation is required to confidently identify true underlying effects; and has been found to underperform in comparison to some automated techniques (Gabard-Durnam et al., 2018). We recommend use of automated artifact rejection procedures to eliminate the potential for researcher bias and allow for the detection of true connection in the EEG., We recognise that automated procedures still need more development (in terms of best techniques) and standardization however, they should be the preferred method of artifact rejection, over manual methods.

To summarise, a detailed account of artifact rejection and controls for each of the types of noise (i.e., electrical, blink, muscle) should always be included. Methodologies proven to remove a significant percentage of artificial noise and maintain more than ~60% of the signal should be used, as suggested by Gabard-Durnam and colleagues. Standardized, automatic processes should be favoured above manual rejection, to increase comparability and decrease variability in cleaning processes and their outcomes, as manual rejection is significantly prone to variability from researcher to researcher and study to study. Lastly, ICA artifact subtraction should be used with caution, and artifact reduction (for example with wICA) should be preferred. Overall, researchers should check the ever-changing literature on the subject and use the suggested updated best practices. For more specifics and how this processing step can be evaluated as part of the checklist, see Figure 1 and the Checklist Scoring in Supplementary Material C.

### 2.2 The effects of VC in EEG-connectivity

VC which refers to the fact that the activity from one source (generated by one brain region or non-brain related EEG artifact) can be picked up by all the EEG electrodes with zero phase delay, as electrical currents are conducted broadly and near-instantaneously through the brain tissue and other matter in the head (Khadem & Hossein-Zadeh, 2014; van Diessen et al., 2015). That is, all electrodes can record some of the same activity at the same time from a single generator of that activity, although at different strengths depending on the distance from the generator of the activity.

It has been suggested that the additive effects of spurious noise and VC in the EEG are particularly influential when connectivity is assessed, most significantly when in scalp-space (Dominguez et al., 2017; Schoffelen & Gross, 2009). Simulation studies have shown that unsynchronised sources that do not possess actual connectivity, when imposed with variability in the strength of their signals (i.e., some sources have high power, some lower) can distort synchronisation measurements at the scalp-level, creating spurious connectivity to varying degrees, depending on the relative powers of the sources (Dominguez et al., 2017; Schoffelen & Gross, 2009). Further still, the dipolar nature of neural activity can create spurious connectivity at electrodes which detect activity from distant sources (Dominguez et al., 2017). Indeed, higher rates of false positives have been reported in scalp-space EEG connectivity (Brunner et al., 2016; Kramer et al., 2008; Lai et al., 2018), and simulation studies have found scalp-based connectivity patterns could be replicated, without any actual source connectivity in their model (Dominguez et al., 2017).

As such, researchers have proposed controlling for the effects of VC by estimating connectivity in source-space data (Schoffelen & Gross, 2009; Cao & Slobounov, 2010). The source-localisation technique statistically separates the EEG into independent signals based on where they have originated, or what “source” is driving a signal, and provides activity values for these independent signals, so that connectivity values can be computed for pairs of these signals (thought to reflect the connectivity between the brain regions represented by these signals; Hassan et al, 2014; Schoffelen & Gross, 2009). Given that the source of a recorded signal (which may be influencing other points) is identified, the effects of instantaneous conduction on a connectivity estimate are minimised (Pascual-Marqui, 2007).

As mentioned in Section 1.1 source-space estimates of connectivity rely on assumptions that are not able to be tested directly in the data and the recovery of the exact time course of sources is lost, thus making the solution of the inverse problem difficult (Dominguez et al., 2017). Further still, source analyses do not completely remove the effects of VC (Dominguez et al., 2017; Schoffelen & Gross, 2009). While current recommendations suggest source-space connectivity to be more robust, the transformation of scalp data to source data requires detailed and complicated explanation. We refer the reader to Jatoi et al. (2014), Khadem and Hossein-Zadeh (2014), and Mahjppry et al. (2017) for in depth explanations of transformations and estimating connectivity in the source-space.

On the other hand, scalp space approaches under specific parameters maybe able to avoid the potential confounds of VC and other spurious noise and may contain fewer assumptions than source approaches making them more robust. Indeed, several studies indicate that the implementation of a statistical test to reject high synchrony with phases around 0 and pi resolves the issue of instantaneous effects with scalp-space connectivity assessments. These instantaneous effects around 0 and pi are suggested to be indicative of volume conducted and non-physiologically plausible connectivity, and their exclusion means that only physiologically plausible connectivity will be included (Dominguez et al., 2017).

Overall, the literature suggests that source-space connectivity can provide several benefits for analysis, especially where the primary research question centres around identifying the shape of connectivity and the understanding of brain region connectivity (Brunner et al., 2016; Lai et al., 2018; Van de Steen et al., 2019). However, scalp-space connectivity assessed using methods to control for VC and with the application of methods such as the surface Laplacian (discussed later), scalp space connectivity can be reliable and valid as it does not make assumptions about or try to model the underlying brain connectivity and sources. Scalp-space analysis might be especially useful in instances where the research question is focused on changes at the individual level across time, or in response to some sort of intervention where the question is not necessarily attempting to identify underlying connections and their shape. Indeed, studies have indicated high test-re-test reliability within scalp-space connectivity measure (Haartsen et al., 2020; Näpflin et al., 2007). The remainder of this guide will focus on optimal connectivity approaches, which can be applied to both source and scalp -space, with adaptations for specific spaces noted.

#### 2.2.1 Controlling for the effects of VC in EEG-connectivity

There are several characteristics of EEG data that can indicate if VC is present in scalp-space data. These include the presence of 0 or pi phase lagged connectivity (electrodes on the opposite sides of a dipole within the brain will receive opposite voltage polarities from that dipole instantaneously with one another, which is reflected by pi lagged connectivity); decreased synchrony strength with increasing distance; only positive correlations between signals in the frequency or time-frequency domains; and positive correlations between connectivity and power at the same frequency (Cohen, 2014). The latter is because a higher amplitude generator will generate higher power measured at electrodes, and because that higher amplitude generator will also reach two electrodes simultaneously through VC, the connectivity will be high also, thus the connectivity and power measures will be correlated.

It is worth noting that VC is always present in the data and is in fact how EEG signals are transmitted from source brain regions to EEG electrodes. As such, VC will always have a potential influence on the amount of EEG-connectivity estimated if it is not controlled for. Thus, in an experiment, not just one but all conditions/groups will be biased by VC. As a result, the comparison between two conditions or two groups may not be adversely affected by VC if both data sets are based on data equally biased by VC. For example, Cohen (2014; Chapter 25) demonstrates that spurious connectivity resulting from bandpass filtering a signal made from random numbers is attenuated when connectivity results are compared across conditions (Cohen, 2014). This feature could further be utilized as a control for spurious connectivity in task-related connectivity, by subtracting a baseline connectivity measure (perhaps during eyes-open resting), to remove the spurious contributions to connectivity which will be common across both the resting and task-related condition (Cohen, 2014). The limitation of this solution, however, is that the effects of VC can vary between conditions/groups depending on external sources, such as noise, and if possible, this noise should be addressed and controlled (via artifact reduction methods that will be described later). However, even when noise reduction methods are implemented, it may be impossible to determine whether the methods have been effective at preventing false positive results related to VC. As such, methods have been developed to mathematically exclude the possibility that VC is responsible for the connectivity measures.

The two most broadly used strategies for mitigating the effects of VC are the use of lagged measures to compute connectivity, and spatial filtering. Lagged connectivity involves the exclusion of the possible effects of zero-phase delay interactions. Zero-phase delay interactions are phase changes that occur at both pairs of electrodes simultaneously, an effect that fits the exact characteristic of VC. This type of connectivity being generated by functional connectivity between brain regions is physiologically implausible for direct connections, given the known (non-zero) duration of transmission times in neural signalling (Cohen, 2015; Michel, 2009), although it is possible for indirect connectivity where two brain regions show instantaneous changes in current due to causal connectivity from a third region connected to both regions (Kovach, 2017).

Indeed, some connectivity analysis methods examine only electrode interactions that have a phase delay (Bastos & Schoffelen, 2016). This is effective at reducing the influence of VC as phase-lag in the frequency domain is equivalent to time-lag between two signals in the temporal domain, and VC cannot explain these delayed interactions as it only produces voltage shifts that occur in both electrodes instantaneously (Bastos & Schoffelen, 2016). Thus, the most probable explanation for the delayed interactions is that the two regions are functionally connected. Several measures of connectivity (such as wPLI, lagged phase synchronisation, and imaginary coherence) employ methods that weight connectivity estimates against zero-phase delays to account for the problem VC. Although different techniques control for the issue in different ways, no techniques account for VC entirely (Cohen, 2015), and there is not yet consensus on the best method to use.

In addition to the zero-phase delay, spatial filters can be used to account for the effects of VC. One such popular filter is the surface Laplacian (also called current source density). The Laplacian works by isolating the distinct activity under each electrode, relative to the closely surrounding electrodes. This approach has been shown to decrease the effects of VC on coherence (Srinivasan et al., 2007) and phase-based connectivity estimates (Cohen, 2015). Dominguez et al. (2017) noted that if each electrode receives contribution only from nearby, local sources, then the analysis of scalp-space data may be justified, especially if the focus is on synchrony between distant regions. Given that the Laplacian removes common data from distant sources from each electrode and leaves only the superficial activity from close to the electrode. The Laplacian is similar to source imaging in the sense that both techniques create virtual channels through linear transformations of the measured EEG signals and provides a useful tool for slap space data analysis. It is important to note however, the Laplacian transform does not eliminate the effects of VC entirely, especially in electrodes close to the ‘seed’ electrode thus the results should be interpreted with caution. Interestingly, using simulated data Cohen (2015) demonstrated that despite the imposed time-lags on the data, the Laplacian was able to identify connectivity and more correctly localise connectivity topographies, again proving a useful tool for scalp-space data.

Overall, the best practice for scalp-based connectivity analysis, as suggested by the existing literature, is to use of estimates that control for VC. Further, studies should employ extra controls for VC such as the surface Laplacian or phase-lags (some connectivity estimates already control for phase-lags, but there is suggestion that implementing a phase-lag in connectivity estimates that do not already employ a phase-lag is useful; Cohen, 2014). Generally, scalp-based measures could perhaps optimally be used where primary research questions centres around changes at the individual level across time, or in response to some sort of intervention. Alternatively, where a deeper understanding of connections and shapes of connections in underlying brain regions is required, source-based measures should be used. For more specifics and how this processing step can be evaluated as part of the checklist, see Figure 1 and the Checklist Scoring in Supplementary Material C.

### 2.3 The effects of the EEG reference

In EEG, voltages are measured as the difference in electrical potential between two electrodes. Usually, this is between electrodes placed across the scalp and a reference site or electrode. Because of the dependence on a reference electrode, activity being received at the reference site influences the signal measured at all other electrodes. Therefore, the reference electrode should ideally be electrically neutral to avoid contamination of the signal(s) of interest. In practice, this is unable to be achieved in the live recording, thus recordings are often re-referenced offline to compute a reference that ideally decreases contamination of the signals. There are several re-referencing techniques for EEG, each with varying levels of bias and effects on the connectivity estimate (Chella et al., 2016; Strahnen et al., 2020; van Diessen et al., 2015).

Many EEG systems use a single reference electrode during online EEG data recording. In this technique, each electrode is referenced against the same, or “common” electrode (Cohen, 2014). However, the use of a single reference electrode to analyse connectivity data is far from ideal (Kayser & Tenke, 2010). If each electrode is referenced against a single common electrode, then pairs of electrodes are exposed to the same signal noise from that reference. Thus, the similarity identified by connectivity analyses between pairs of electrodes might be produced by this reference signal commonality, rather than synchrony of the activity generated by the cortex (Chella et al., 2016; Zaveri et al., 2000). Bastos and Schoffelen (2016) further demonstrated this in a simulated dataset, where no real coupling was present, yet artificial connectivity was observed between two sources due to the common reference.

Other research has used mastoid (or ‘earlobe’) referencing - this is where the signals are averaged to recordings from the back of the left and right ear (Qin et al., 2010). It is sometimes assumed that the data recorded from the mastoids or earlobes is not brain-related but is still close enough to other channels to pick up similar non-brain noise in the data. This therefore enables this reference technique to remove noise from the signal detected at each electrode, but not reduce brain-related contributions to the data (Chella et al., 2016). Thus, the technique is proposed to remove noise that is thought to equally affect all the channels. While mastoid referencing was found to perform better than the use of a single common reference, Chella et al. (2016) demonstrated that it still distorted the EEG signal and the resulting connectivity estimate.

These distortions to EEG connectivity as a result of mastoid referencing could be because it is not necessarily true that activity at the mastoids is not brain related. For example, research into mismatch negativity (MMN), an EEG event related potential (ERP) has noted that MMN components and the differences in MMN between populations can be best distinguished at mastoid sites (Hirose et al., 2014; Sussman et al., 2015). This suggests that mastoid sites are not in fact ‘neutral’ and that using mastoid referencing is confounded by the effects of underlying brain activity on the mastoid reference, which can inflate connectivity estimates without true connectivity between the brain regions being assessed.

The problems with the single reference and mastoid/earlobe referencing techniques discussed so far have been addressed through many means, one of which includes employing common average re-referencing (CAR). CAR involves calculation of the average activity of all recording electrodes, and then subtracting this average from each electrode (Lei & Liao, 2017; van Diessen et al., 2015, for other methods see Lepage, Kramer, & Chu, 2014; Tenke & Kayser, 2015). CAR generates less EEG signal distortions and the resulting connectivity estimate than the single reference and the mastoid/earlobe reference (Chella et al., 2016). Research using simulated EEG data has demonstrated the CAR can be improved by adapting a robust maximum likelihood estimator (see Lepage, Kramer, & Chu, 2014 for more), dubbed the robust CAR (rCAR) which performed better than the other techniques.

The Reference Estimation Standardisation Technique (REST) is another method designed to address the issues with CAR. The REST technique mathematically reconstructs the data using an equivalent distributed source model to re-reference the data offline against a point at infinity (for more see source paper by Yao, 2001). This is believed to be the ideal 'neutral' potential that does not use body surface points which can negatively influence the data (Yao, 2001). Chella et al. (2016) compared the differing effects of reference choice on both simulated and real EEG data and found that the REST re-referencing technique led to the least distortion of the connectivity measure. This is a similar finding to Qin et al. (2010) who found the infinity reference to outperform other referencing techniques (except rCAR) when estimating coherence. However, Lepage et al. (2014) included the REST technique as a comparison when assessing the rCAR re-referencing approach and suggested that rCAR outperformed REST. Other referencing techniques inflated the connectivity values obtained in the following order (from most inflation to least): Cz, mastoid referencing, and CAR (Chella et al., 2016). Thus, so far it seems that REST and CAR (particularly rCAR) were the two better re-referencing choices. Interestingly, REST re-referencing has been found to be a robust and reliable technique in other EEG modalities, including task-related and ERP measures (see Mahajan et al., 2017).

It should be noted that CAR and REST are not entirely free of limitations. The primary limitations are due to insufficient scalp coverage (Lei & Liao, 2017; van Diessen et al., 2015; Yao, 2001). This is especially important for the performance of REST. Chella et al. (2016) demonstrated that higher-density EEG systems were better able to reconstruct network patterns of activity in simulated data. That is, when fewer channels were used for the REST re-reference, the chances of a false positive result in later connectivity estimates are increased. Therefore, no less than 32 channels should be used with the REST re-referencing technique when considering connectivity analyses (Chella et al., 2016).

Furthermore, we suggest this low limit should also be applied for CAR and rCAR, as both techniques are based on the idea that more electrodes represent a good sampling of the brain thus, the average potential is zero or as close to it as possible. It is unlikely that a good sampling of the head can be achieved with less than 32 electrodes with CAR, and even more electrodes are needed for rCAR, as the robust estimation method is based on discarding information coming from some sensors.

Lastly, the spatial filter, surface Laplacian (discussed above) is a reference-free technique which has been demonstrated to increase the accuracy of functional connectivity estimates, specifically phase-based (e.g., inter-site phase clustering [ISPC or just ‘phase synchronization’] and wPLI; Cohen, 2015; Kayser & Tenke, 2015). Indeed, the surface Laplacian has been found to be superior to CAR at VC artifact attenuation in connectivity studies (Cohen, 2015; Srinivasan et al., 2007). Given several claims that false positives are increased when assessing connectivity in the scalp-space (Brunner et al., 2016; Lai et al., 2018), and the is growing suggestion that scalp-space measures of connectivity are not ideal, the surface Laplacian can mitigate these potential confounds, and possibly make scalp-space estimates more robust given that it accounts for VC, assesses current flow, and is much simpler and based on fewer assumptions than source-localisation techniques (Cohen, 2015).

Overall, the REST, rCAR and the surface Laplacian (specifically for scalp-space connectivity) seem to be the most robust referencing techniques, producing the least amount of distortions to the EEG-connectivity estimates. Ideally, reliable results should be robust against variation in methods, so a desirable confirmation step for researchers would be to confirm their results with an alternative (but still robust) re-referencing method (which could be reported in the supplementary materials or just as a note that the results replicated with multiple re-referencing approaches). For more specifics and how this processing step can be evaluated as part of the checklist, see Figure 1 and the Checklist Scoring in Supplementary Material C.

### 2.4 The effect of epoch length and number

Connectivity estimates can be significantly impacted by the length of data segments or ‘epochs’ that are examined, and the number of data segments or points included (Chu et al., 2012; Haartsen et al., 2020). If the EEG epochs being analysed are shorter than the interactions in activity between the recording electrodes, results can be biased due to unreliable estimation of the frequency spectra (this is especially influential for slower, low frequencies; Bakhshayesh et al., 2019; Fraschini et al., 2016). Shorter epochs may not allow for the whole interaction to be assessed, producing a connectivity estimate that suggests no underlying connectivity when there may in fact be connectivity.

Overall, research does suggests that longer epochs demonstrate more stable patterns of connectivity (Chu et al., 2012). Fraschini et al. (2012) were able to differentiate more accurately and shown that at least 4 second epoch show patterns of stability for varying connectivity measures, however for the two measures used 6 seconds was best for AEC 12 second for PLI. Unfortunately, the authors did not differentiate between epoch length according to frequency, nor did they transform the data into frequencies, or assess how the number of included epochs might affect the analysis.

Furthermore, in addition to the importance of epoch length, it has been demonstrated that connectivity estimates based on lower numbers of epochs can bias the results towards false positives (Bastos & Schoffelen, 2016). This is especially the case where two or more conditions are compared, and the number of epochs assessed in different conditions is uneven. Bastos and Schoffelen (2016) demonstrated that basing connectivity estimates on less than 100 epochs on average resulted in higher estimates of connectivity (they specifically assessed coherence, Granger causality, and phase-locking value). The authors suggest that in general, a smaller number of epochs increased the bias in the connectivity estimate. To add to the complexity of these considerations, it may also be that the interaction between number and length of epochs can adversely affect connectivity measures.

A recently published paper assessing infant EEG test-retest reliability may help shed some light on the interaction between number and length of epochs. Haartsen et al. (2020) assessed alpha specific connectivity using the phase lag index (PLI) and debiased weighted (dbW)PLI. It was noted that reliability of whole brain dbWPLI was higher across many short epochs (50 x 2 seconds) but for PLI, reliability was higher across fewer longer epochs (20 x 4 seconds). Overall, reliability of the whole brain connectivity metrics was higher than for connectivity using network metrics (discussed below, see Supplementary Material A for definition). While this study took place in infants, test-reset reliability in the study was high and similar results have been found in adults with the dbWPLI where many short epochs were utilised (although with little comparison to other conditions; Kuntzelman & Miskovic, 2017; Vinck et al., 2011).

The interaction between epoch number and length is even more important to understand in studies where task-related EEG-connectivity is assessed because the use of longer epochs may not be possible. Depending on the task, stimuli may be presented to the participant with less than 2 seconds separating each stimulus presentation. If longer epochs *are* better (as suggested by the literature) and apply that standard to task EEG conditions, this could mean the possibility of assessing connectivity across multiple, different brain processes as connectivity may change on relatively short time scales during cognitive processing. Thus, a specific brain process of interest, which may take place during only a short time window following task stimuli may not be assessed. In this case, it may be necessary to adjust the methods and consider adding more epochs. Thus, we recommend substituting more trials for longer epochs, based on existing research we tentatively recommend no fewer than 50 trials for epoch less than 2 s (Haartsen et al., 2020; Kuntzelman & Miskovic, 2017; Vinck et al., 2011).

The use of more epochs in connectivity analyses may further allow for a reduction in spurious connectivity even for shorter epoch lengths. This point has been demonstrated in resting-state studies where longer epochs of artifact free data were not available; research by Haartsen et al. (2020) showed that a larger number of shorter epochs, provided more reliability in assessing connectivity. However, these results are restricted to the alpha frequency, and it may be that connectivity in different frequencies interacts differently with epoch length.

It could be that shorter epoch periods may be more reliable for faster frequencies (Haartsen et al., 2020) and longer windows for lower frequencies (Chu et al., 2012). This arises from the circumstance that for faster frequencies a shorter time window can capture enough oscillatory cycles for the purposes of comparison, as opposed to lower frequencies for which fewer oscillatory cycles maybe captured in shorter time windows (Cohen, 2014). Overall, it is suggested that studies assessing both resting-state and task-related connectivity provide specific justification for epoch length being assessed. This means taking into consideration the underlying brain processes, the frequency of interest, the connectivity estimate being utilised, and for task-related studies, the parameters of the specific cognitive task being used. In general, and until more robust research is produced, the 4-6 second guidelines seem optimal for phase-based connectivity measures.

To summarise, epochs of lengths longer than 4 seconds should be used to identify robust and consistent EEG-connectivity. However, longer epochs are not always possible in task-related methodologies where a brief burst of connectivity may last for 500 ms, which may not be detected if 6 second epochs are used. Although current research has not provided methods to circumvent this issue, there are less favourable and more favourable approaches. If shorter epochs are used, we recommend no fewer than 50 epochs be analysed We also recommend including a clear rationale based on past research literature, ideally, one based on underlying brain processes relating to the condition, disorder, or cognitive task being assessed (i.e., focused on time-period when modulations of connectivity are expected within the task).

Finally, where initial studies seem to suggest longer epochs to be more reliable for assessing lower frequencies and shorter epochs for higher frequencies, no specific recommendations are made as part of the checklist, this due to limited literature on the specific differences. For more specific details and information about how this processing step can be included in the checklist, see Figure 1 and the Checklist Scoring in Supplementary Material C.

**2.5 Network theory and topography in EEG-connectivity**

An important option for performing EEG-connectivity analysis is topographical mapping. Network theory is an applied form of graph theory which is a framework for the theoretical means of structuring/modelling connections to evaluate connectivity patterns between brain areas, ‘nodes’, or electrodes (in the case of EEG; Sporns, 2011). Non-network approaches look at specific pairs of electrodes that show connectivity between the two electrodes: they are not evaluated with respect to how that pair is connected to a broader network. Conversely, network metrics are effective and more robust at identifying connections in larger sets and sequences of electrodes, known as multivariate analysis.

However, multivariate analyses in network metrics are very complex, and EEG studies generally obtain bivariate connections and apply them to network metrics using a variety of techniques. Within these, the matrices of connectivity strengths for each pair of electrodes are often further simplified by transforming graded matrices with continuous values into binary values (i.e., 1 = connectivity present; 0 = connectivity absent; Peeples & Roberts, 2013). There are several ways in which EEG-connectivity can be binarized, including Cluster Span Threshold (CST; Smith et al., 2015), and Minimum Spanning Trees (MST; Stam et al., 2014); these techniques are used to determine which connections belong in the network metrics and which do not. That is, the different binarization methods incorporate varying threshold settings to construct synchronised networks based mostly on the strongest connections, obtained from the time-varying oscillations of different brain regions (Bassett & Bullmore, 2006).

Thresholding in graph metrics is not used to evaluate statistical significance of the connections but to set the data up for further evaluation and analyses (Cohen, 2014). How the threshold is applied depends on the purpose of the data analysis. Generally, however, techniques that incorporate arbitrary and subjective threshold setting should be avoided as they can cause problems for data analysis. That is, studies that set arbitrary thresholds that can vary between studies can end up producing different results, even when using the same measure (for an example, see Sun et al., 2019). If certain connections do not meet the threshold set by a particular study, they will not be included in the network. Whereas, if in another study the threshold is set lower or higher, their results may show more, or less significant connections, respectively, which will influence the result and the reliability of the comparison between the two studies.

Further, given the fact that the threshold and its values are subject to change depending on the distribution of the data, a threshold that is based on the data itself but independent of comparisons between conditions is important. We strongly recommended defining the threshold not numerically but based on density (the number of connections one has, compared to the number of total possible connections they *could* have), as the density of the graph strongly affects the results (see Sanz-Arigita et al., 2010; van Wijk et al., 2010). Techniques such as the MST (Stam et al., 2014) and efficiency cost optimization (De Vico Fallani et al., 2017) overcome the problem of network density. Using data-driven analyses that apply “weighting” to connections may be a more acceptable alternative; this weighting is based on the connection’s respective strengths (De Vicco Fallani et al., 2014). Thus, stronger connections hold more “weight” in the network whereas weaker connections hold less, but all connections contribute to the final analysis. Where non-arbitrary threshold setting is unavailable, it is recommended that several thresholds are assessed and reported (Rubinov & Sporns, 2010). All tested thresholds should be reported in publication and ideally pre-registered, to avoid the potential for ‘fishing’ for positive results.

Ultimately, networks are thresholded and binarized with the aim of setting up the data for subsequent analyses (such as connectivity degree, node clustering, or small world networks). While network analyses provide insight into the topography of connections and can minimise some of the statistical limitations, such as the problem of control for multiple comparisons (given that assessment is performed at the level of the network rather than for each individual connection), network analyses only consider the transformed binary relations. That is, network analyses may not characterise the specific network properties accurately, given that connectivity strength is a continuous measure, and network analyses only consider binary relationships between electrodes / regions (Peeples & Roberts, 2013). Therefore, as with most processing and analysis choices, it is important to consider the aim, hypothesis, and specific research question(s) before deciding whether to incorporate network theory.

To summarise, when using and including network metrics in research studies, a clear plan and rationale needs to be presented before data analysis. Networking threshold should not be set arbitrarily, rather objective model-driven thresholds should be employed. Alternatively, where researchers do not wish to disregard lower-strength connections, weighted networks maybe employed to give an understanding of whole networks. For more specifics and how this processing step can be evaluated as part of the checklist, see Figure 1 and the Checklist Scoring in Supplementary Material C.

**2.6 Statistical considerations for EEG-connectivity assessment**

A primary consideration in connectivity statistics is the multiple comparison issue. Given that the number of electrode pairs that could be included in comparisons increases as the number of electrodes that are available for analysis increases, there is the potential for electrode pairs to assess connectivity within numbering >1000 for caps with >64 electrodes. If statistical comparisons are conducted between experimental groups for every pair of electrodes, and multiple comparisons are not controlled for, this number of comparisons can almost guarantee false positives. Despite this risk, some studies neglect to control for multiple comparisons at all, while other studies fail to completely account for multiple comparisons.

Common approaches to controlling for multiple comparisons include the Bonferroni correction and the False Discovery Rate (FDR). However, the adjustment of the *p*-values in for the two methods can greatly reduce the statistical power, produce too stringent values and potentially hide actual effects; the tests cannot adequately account for the real number of comparisons in almost any functional connectivity analysis; and the Bonferroni correction is too strict in its assumption of independent statistical tests (Cohen, 2014; Zalesky et al., 2010). EEG data is correlated and multidimensional thus, the values of functional connectivity in different pairs of electrodes are not independent.

A common and recommended approach in accounting for the problem of mass-univariate comparison testing in connectivity analyses, is the incorporation of non-parametric permutation testing. Non-parametric permutation testing does not rely on assumptions about the theoretical underlying distribution of test statistics (as do parametric tests) rather, on the distribution created from the available data. This framework is most effective in non-normally distributed data (a characteristic that EEG metrics commonly possess). Permutation testing is not a control for multiple comparisons, but it allows for easy incorporation of corrections for multiple comparisons that provide *p*-value thresholds sensitive for correlated multidimensional data, and correct the value based on information in the results rather than the number of tests.

The most common method of multiple comparison control available for nonparametric permutation testing is correcting by using the cluster size of connections between neighbouring pairs of electrodes to determine the threshold. The detailed methodology of these two techniques is beyond the scope of this paper – for a starter in these EEG connectivity analyses, please see Cohen (2014). It should be noted when nonparametric permutation tests are combined with graph theory and thresholding (described above), they can help to avoid modelling issues and problems derived from deviations from normality (see Maris & Oostenveld, 2007). One example of a nonparametric procedure, called the network-based statistic (NBS), was developed specifically to offer greater power than what is possible when independently correcting the *p*-value, with the addition of the underlying assumption that the connections of interest form components. Here, the NBS helps in detecting influences of interconnected subnetworks that other approaches cannot thus, NBS controls for family-wise error rates when mass-univariate testing is performed at every connection comprising the graph or connectivity metric [see Zalesky et al. (2010) specific on the NBS methodology and Han el at. (2013) for NBS and cluster size combined methods].

While NBS and cluster-based corrections are commonly used and feasible, the thresholds used to define the sets of links are often not data-driven, thus, based on pre-defined, subjective choices which can produce differing results (Langer et al., 2013; Maris & Oostenveld, 2007). We recommend data-driven method of cluster thresholding, alternatively, there is some suggestion these methods can be uniquely defined from the spatial frequency of the data (see Flanding & Friston, 2019). By way of efficacy and applicability in connectivity analysis the authors recommend using non-parametric permutation testing when: data is non-normally distributed; graph theory-based analyses are being assessed; analysis is data-driven and/or exploratory rather than hypothesis-driven; and when multiple network connections (i.e., mass-univariate tests) are being assessed rather than *global* network measures (Cohen, 2014; Zalesky, 2010).

Overall, when applying nonparametric permutation tests, it is recommended researchers are specific about determining and reporting: the property of the data being shuffled in the nonparametric permutation test, and that the shuffling is appropriate to the focus of the research question (Theiler et al., 1992); the number of iterations performed; the creation of the *p*-value (or the *p*-value threshold used in null-hypothesis for obtaining clusters); and the specifics of the correction for multiple comparisons (Cohen, 2014; Maris & Oostenveld, 2007).

To summarise, if multiple comparisons are controlled for by limiting the analysis to averaged whole-brain connectivity or a small number of specific pairs, using the Bonferroni or FDR methods is appropriate. However, where mass-univariate testing is the key statistical model, non-parametric permutation tests and cluster-statistics may be best. For more specifics and how this processing step can be evaluated as part of the checklist, see Figure 1 and the Checklist Scoring in Supplementary Material C.

### 2.6 Sample size considerations

As a final consideration, Larson and Carbine (2017) demonstrated that zero studies out of 100 high-impact, randomly selected EEG studies reported statistical power calculations for participant sample size selection. A priori calculation of the sample size required for sufficient statistical power is important to ensure studies are adequately powered to detect the effect of interest. It is also important for meta-analytical inspection of research (see Thorlund et al., 2011). Yet, many studies commence without proper a priori calculations for the required sample size.

Given the existing discrepancies in sample size effects, selection and analysis, no clear criteria or recommendations can be made. However, the authors encourage researchers to increase attention to the reporting and performing of a priori sample size calculations (as outlined in Supplementary Material B), as well as the presentation of this information and the associated statistics, as this step will greatly increase scientific rigor in future EEG-connectivity research. For some suggestion and further discussion see Supplementary Material B: On Sample Size. For the present checklist, the authors included a simple two option item assessing if sample size was based on any statistical considerations or not, as outlined above, see Figure 1 and the Checklist Scoring in Supplementary Material C.

### 2.7 Assessing connectivity with EEG: Final considerations and a summary

It is important to be aware that there is currently no “best approach” in analysing connectivity for all data. Bakhshayesh et al. (2019) compared 26 measures of connectivity on generated, synthetic data. Their results indicated that noise level, stationarity of data, the number of samples, the duration of EEG data available for analysis, and the number of channels greatly affects the connectivity estimate. Thus, when choosing which analysis methods to use, care needs to be taken and method selection should be based on considerations of these, and the analyses that have been shown to work best for that type of data should be used.

Interestingly, a study in rodents assessing local field potentials noted a considerable lack of functional redundancy and little covariance across multiple measures of connectivity estimates (Strahnen et al., 2020). Evidence for this non-redundancy has been observed in human EEG where network measures, based on source-level EEG with a dependency on arbitrary network metrics, were employed (Fraschini et al., 2020). Thus, suggesting that future EEG-connectivity studies may benefit from basing conclusions on more than one metric and making the analysis approach explicit. While this is not a requirement for the checklist presented in this article, we encourage readers to seriously consider and utilise their recommendations.

To summarise: robust common average re-referencing or REST referencing are the two best methodologies for correct connectivity identification (potentially with the addition of Surface Laplacian transforms). Further, epochs that are longer than 4 seconds provide more accurate connectivity estimates, and epochs longer than 6 seconds provide optimal connectivity estimates. Additionally, the use of 100 or more sample epochs (which are equal in number across all conditions) reduces connectivity estimate biases. Controlling for the effects of VC (through phase-based measures, measure employing phase-lags, or source localisation techniques) is another important step for accurate connectivity estimation. Artifact removal methodology should be described in detail accounting for eye blinks, muscle movements, and electrical noise, naming any extra analyses (i.e., ICA) or pipelines/toolkits applied to the data cleaning process. Lastly, controlling for multiple comparisons is essential, and employing objective, model-based network metrics or utilising multiple connectivity estimates may increase the reliability and validity of the connectivity estimate.

## 3 EEG-connectivity Research Quality Checklist

With the issues and considerations regarding the interpretation and reporting of EEG-connectivity results identified and explained, we sought to develop a novel checklist to allow for quality assessment and comparison of EEG-connectivity studies. Ideally, the checklist will inform future research and ensure high quality research design for EEG-connectivity studies. Unfortunately, the evidence base available cannot provide specific recommendations for all steps due to the range of potential analysis techniques and lack of direct companions to demonstrate superiority of one over the other, the checklist provides specific recommendations as to which approaches should be considered sufficient to produce high-quality studies of EEG connectivity. Thus, specific recommendations have been provided where possible and a range of options where the evidence-base does not make clear a single best solution. As a result, the ‘study quality’ checklist in Figure 1 is proposed for the assessment and interpretation of EEG-connectivity studies, each item on the checklist is annotated to provide clear guidelines as to what to assess.

This checklist can serve multiple purposes including: 1) guiding what critical information authors should be implementing in their studies and reporting; 2) indicating the information peer reviewers and journal editors should be assessing and/or requesting; and 3) suggesting the information that readers of EEG-connectivity research should be critically evaluating. Therefore, the goal of this checklist is to present specific and key methodological guidelines for consideration, prior to commencing an EEG study or analysis, when publishing results, and when interpreting the results of an EEG-connectivity analysis/study.

The checklist is by no means intended to limit researchers. Rather, the aim is to draw attention to the potential for false positives when evaluating EEG-connectivity research, and areas for potential improvement for future research. Additionally, this checklist is specifically tailored for studies and researchers assessing EEG-connectivity, not EEG in general (for existing guidelines on general EEG processes, see Keil et al., 2014).

**Figure 1.**

*EEG-connectivity Study Checklist.*

| | 0 | 0.5 | 1 |
|---|---|---|---|
| 1. Re- referencing technique: | Single, mastoid / ear, nose, or not presented ☐ | CAR ☐ | REST, rCAR, or Laplacian ☐ |
| 2. Epoch length: | <3 s ☐ | 4-6 s ☐ | >6 ☐ |
| 3. Number of sample epochs: | <50 ☐ | 50-100 ☐ | >100 ☐ |
| 4. Artefact rejection technique: (for specifics see "Checklist Scoring Specifics" section) | None ☐ | Noisy epochs, or channels rejected ☐ | All types of artefact addressed ☐ |
| 5. Control for volume conduction: (for specifics see "Checklist Scoring Specifics" section) | None ☐ | Lag, weighted, source, or Laplacian ☐ | (Lag OR weighted) & (source-space OR Laplacian) ☐ |
| 6a. Control for multiple comparisons: (for specifics see "Checklist Scoring Specifics" section) | No post hoc) ☐ | Invalid post hoc control OR p-value = 0.01 ☐ | Valid post hoc control ☐ |
| 6b. If network / cluster -based Statistics use this guide (and do not code 6a) | | Arbitrary thresholding / not model-driven ☐ | Model OR data -driven threshold or weighted ☐ |
| 7. Sample size estimation and consideration | No considerations ☐ | Some consideration (i.e., *N* obtained from published literature) ☐ | Statistical consideration (i.e., a priori *N* calculation) ☐ |
| **Row scores:**<br>**Total score & QR:** | | | |

*Note.* 0 = not recommended for use; 0.5 = not optimal; and 1 = optimal for use. CAR = common average reference; rCAR = robust CAR.

The complete, detailed scoring guidelines for the checklist are included in the Supplementary Material C: Checklist Scoring Guidelines. The checklist can be scored according to the criteria outlined in Table 1, with studies falling into one of 3 categories: high, moderate, or low quality.

**Table 1.**

*EEG-connectivity Scoring and Level of Study Quality.*

| Quality level | Description | Criteria that need to be met |
|---|---|---|
| 1 | High quality | Quality checklist score between 5+ |
| 2 | Moderate quality | Quality checklist score between 3.5 and 4.5 |
| 3 | Low quality | Quality checklist score between 0 and 3 |

It is important to note that the quantitative scores should be solely used for the purpose of making the checklist easier to score into the qualitative framework below (Table 1). In this sense, “0” corresponds to “not recommended for use,” “0.5” to “can be used,” and “1” to “optimal for use.” The qualitative measure does not assume that for example, REST referencing is three times better than average and two times better than mastoid, nor does it imply that mastoid should not ever be used. It also does not mean to imply that the use of mastoid re-referencing is equivalently vulnerable to false positives when compared to no multiple comparison controls. In certain circumstances, the lack of multiple comparison controls could almost guarantee false positives, while the use of mastoid re-referencing is likely to result in a small incremental increase in the risk of false positives, as discussed previously. However, to raise awareness and promote best, standardised practice amongst researchers, methodologies with “0” or “not recommended for use” should be discouraged as those approaches increase the likelihood of false positive results.

As the field of EEG-connectivity continues to develop and more studies are published, other criteria may be considered and added to the checklist. Consideration might also be given to points such as reporting of *p* values for each electrode pair comparison, or perhaps the reporting of the means and standard deviations or effect sizes for connections between individual electrodes (these values or checklist parameters could be reported as supplementary material). This approach would allow for meta-analytical methods to be developed similar to the activation likelihood estimation meta-analytical approach in the field of fMRI.

The checklist and its recommendations are made to make the results of the field more synthesisable. The use of the checklist may also enable better quantification of EEG-connectivity study quality, for the purposes of a meta-analysis, which is not possible with the field as it stands. However, should the checklist be used in subsequent meta-analyses, the results of the meta-analysis could more reliably inform us about several connectivity related processes as a result of an improved ability to assess the quality of the included studies. For example, we may be better able to answer questions like: “how likely it is that connections between specific regions or in specific bands (or both), are impaired for different neurological or mental health disorders?” Or “how robust is the evidence regarding connectivity increases related to specific cognitive functions, and what is their relation to changes in connectivity?”. Such connectivity related questions are valuable and deserve rigorous evaluation. It is our hope that the identification, explanation, and assessment of EEG-connectivity research components will assist the field in answering important questions such as these.

**Authorship Confirmation Statement**

All who meet authorship criteria are listed as authors, and all certify that they have participated sufficiently in the work to take public responsibility for the content. With AM contributing to conceptualization, design, writing – preparation, creation, writing – editing, and revision; NWB to conceptualization, design, writing – editing, and revision; and SEH, FVR, and PBF contribution to the writing – editing, and revision.

**Disclosure, Declaration of Interest and Funding**

Authors AM, NWB, and SEH report no financial or potential conflicts of interest.

PBF is supported by a NHMRC Practitioner Fellowship (1078567). PBF has received equipment for research from MagVenture A/S, Medtronic Ltd, Neuronetics and Brainsway Ltd and funding for research from Neuronetics. He is on scientific advisory boards for Bionomics Ltd and LivaNova and is a founder of TMS Clinics Australia.

FVR receives research support from CIHR, Brain Canada, Michael Smith Foundation for Health Research, Vancouver Coastal Health Research Institute, and in-kind equipment support for this investigator-initiated trial from MagVenture. He has received honoraria for participation in advisory board for Janssen.

**Supplementary Material A:**

**Key term definitions**

The following section presents a table of key definitions (A–Z) relating to the manuscript 'EEG-connectivity: A fundamental guide and checklist for optimal study design and evaluation' all definitions are orientated and presented in the context of the brain imaging tool, Electroencephalography (EEG).

| Term | Definition |
|---|---|
| Anti-phase | In EEG sinusoidal waves, anti-phase refers to when two signals are inversely correlated (i.e., a high positive voltage is present in one electrode and a high negative voltage is present in another electrode). |
| Bandpass filtering | Filtering oscillations in the EEG signal to limit the oscillation bandwidth present in the data, which prevents the interference of frequencies outside of the desired band (i.e., non-brain signal frequencies) Most commonly, slow frequencies of less that 0.1 Hz and high frequencies above 40-50 Hz are filtered out in EEG. |
| Bivariate analyses | Comparison or analysis of two variables. Thus, within EEG connectivity - electrode-to-electrode or brain region source-to-source relationships. |
| Connectivity (general) | Synchronized activity of brain regions within connected networks quantified based on the |
| | simultaneous patterns of some sort of activity (i.e., haemodynamic, or electromagnetic) in separate brain regions across time |
| Dipole | At its simplest, a circular electrical charge with an equal magnitude in power, but opposite charges (positive and positive) at opposite ends, dipoles can be larger or smaller in the distances they cover. |
| Effective connectivity | Brain regions showing a relationship based on some sort of activity that is unidirectional in nature, it attempts to determine the casual flow of information from one point to another (i.e., if activity in one brain region precedes activity in another region in time; Friston, 2011) |
| Frequency | Macroscopic neural oscillations that occur at varying times per second thus can be alpha (i.e., 8-12 oscillations per second or 8-12 Hertz [Hz]), theta phase (i.e., 4-8 Hz), beta phase (i.e., 13-30 Hz), etc. |
| Functional connectivity | Also brain regions showing a relationship based on some sort of activity but bi-directional in nature and cannot determine causation (i.e., are two brain regions sharing common activity, suggesting they are connected; Friston, 2011). |
| In-phase | When two recorded signals oscillate at the same rhythm (i.e., when the crests of a signals/wave pass the same points at the same time). |
| Multivariate analyses | Comparison or analysis of more than two variables, electrode-to-electrode-to-electrode or source-to-source-to-source (or more) comparisons. |
| Network structures or network modelling or network theory | Network theory is an applied form of graph theory which is a framework for the theoretical means of structuring/modelling connections to evaluate connectivity patterns between brain areas, 'nodes', or electrodes (Sporns, 2011). Non-network approaches look at specific pairs of electrodes that show connectivity between themselves: they are not evaluated with respect to how that pair is connected to a broader network. |
| Noise / artifact | Signals recorded by the EEG that are not generated by the brain. These can be psychological noise such as eye blinks and, muscle or head movements, or electrical/line noise. |
| Oscillations | Voltages shifting from negative to positive and back again, multiple times per second, usually in a regular rhythm or phase |
| Over- / Across- time connectivity | Connectivity measured between two or more electrodes/brain regions across time. |
| Over - / Across- trial connectivity | Connectivity measured between the phase of oscillations and the presentation of stimuli over multiple presentation of that stimuli, within a single electrode. |
| Phase | Phase specifies the timing or location of a (given) point within a wave cycle of a repetitive waveform. Thus, the phase refers to the characteristics of a wave. |
| Power | The magnitude of oscillatory cycles within a specific frequency from peak to trough. |
| Referencing | An electrode signal or signals against which the potential of another single electrode is compared to determine differences in potentials. |
| Resting-state recordings | EEG activity recorded while the research participant is not engaging in any goal- or task- orientated cognitions, rather is sitting alert but restfully. |
| Signal- / Scalp- space | EEG activity assessed at the level of the EEG electrodes and comparing activity between EEG electrodes. |
| Source | An origin site for the recorded EEG activity, responsible for or driving the observed EEG activity. |
| Source-space | EEG activity transformed to estimate the location and distribution of the signals, to identify ‘sources.’ When this is achieved, this is referred to as solving ‘The Inverse Problem’ (defined below). |
| Structural connectivity* | Based on anatomical connections between regions (i.e., white matter tracts), cannot be measured by EEG. |
| | *This type of connectivity is not mentioned or included in the manuscript, since it is not directly relevant to assessing EEG connectivity, however it is presented here for those seeking distinction. |
| Task- or event- related regions | EEG activity recorded while the research participant is performing a specific cognitive task. |
| The Inverse Problem | The process of calculating and thus making inferences as to the position of the current sources of signals recorded by EEG electrodes (Grech et al., 2008) |
| Topographic maps (in EEG) | Topographic maps are a detailed record of electrodes and electrode placement, giving positions and features of the EEG signal. |
| Volume conduction / Field spread | A description of the phenomena whereby electrical voltages spread throughout a medium. Within EEG connectivity, this is relevant for the fact that the activity from one source, generated by one brain region or non-brain related EEG artifact, can be picked up by all the EEG electrodes without delay (i.e., zero phase delay). Thus, all electrodes can record some of the same activity at the same time from the single generator of that activity, although at different strengths depending on the distance from the generator of the activity (Khadem & Hossein-Zadeh, 2014; van Diessen et al., 2015). Volume |
| | conduction thus creates artificial synchrony between signals and affects functional connectivity estimates for measures which do not control for these near-instantaneous effects. |

## Supplementary Material B:

## On Sample Size

As discussed, Larson and Carbine (2017) demonstrated that zero studies out of 100, randomly selected EEG studies from high impact journals reported statistical power calculations for participant sample size selection. A priori calculation of the sample size required for sufficient statistical power is important to ensure studies are adequately powered to detect the effect of interest. It is also important for meta-analytical inspection of research (see Thorlund et al., 2011). Yet, many studies commence without proper a priori calculations for the required sample size. Further still, there is the argument that most studies in neuroscience are severely underpowered and that larger sample sizes are needed to produce more reliable results (Friston, 2013; Larson & Carbine, 2007).

Interestingly, Friston (2012) presented an analysis of effect size in classical inference and demonstrated that a sample size of more than 50 individuals resulted in exposure to trivial effects and loss of integrity; this is because the null hypothesis will be rejected with probability of one (i.e., it *will* be rejected) in the presence of a trivial effects, as per the fallacy of classical inference (for more see Friston 2012; Senn, 2001). Overall, an optimal sample size between 16 and 32 individuals has been suggested for neuroscience studies (Friston, 2012). However, this is should *not* be seen as a recommendation for employing smaller sample sizes, only that smaller sample sizes can still produce reliable results. Without stronger evidence specific to connectivity, the authors suggest using no less than 30 participants per group to obtain an appropriately representative and generalisable sample. However, where possible, more data should be collected, and appropriate controls and sound statistical knowledge should be used to address the problem of trivial effects.

Given the existing discrepancies in sample size effects, selection and analysis, no clear criteria or recommendations can be made. Thus, sample size was not added as an item in this checklist assessing key components of EEG-connectivity research studies. However, we recommend increasing attention to the reporting and performing of a priori sample size calculations, as well as the presentation of this information and the associated statistics, as this step will greatly increase scientific rigor in future EEG-connectivity research. Sample size was not included as part of the checklist however, the authors encourage researchers to undertake a priori sample calculations and take care to protect against inferences on trivial effects in larger sample sizes, (i.e., using protected inference), and additionally that *p*-values be considered *with* point estimates and standard errors, confidence intervals, or likelihood functions.

**Supplementary Material C:**

**Checklist Scoring Guidelines**

For Item 1, if the method of referencing is not mentioned, the single common reference, or mastoid referencing is used, then a score of 0 should be allocated. If average re-reference is employed, then a score of 0.5 should be allocated. Lastly, if surface Laplacian, REST, rCAR, or another demonstrably robust method for connectivity re-referencing is employed then a score of 1 should be allocated. It should be noted that for the impact of reference electrode choice on the probability that results are confounded could depend on whether data is being assessed for comparisons between groups or conditions. For example, the mastoid reference may bias the estimate of the connectivity or produce a possible change on spatial resolution (i.e., blurring or spread of the estimation). This effect can be similar for both conditions/groups, so the statistical results might not be affected by use of the mastoid reference if between group / condition comparisons are being made. The likelihood of a false positive being detected is therefore greater if directly looking at the connectivity values, rather than differences between conditions/groups of connectivity values. However, if one group/condition is more likely to have noisy mastoids, then it could lead to increased false positives, or if the noise is not specific to just one group/condition, then the noise could lead to reduced signal-to-noise ratio, and potentially false negatives. Therefore, the authors take the conservative view and make the above recommendations for item 1.

For Item 2, if epoch length is not mentioned or the epochs are less than 3 seconds, then a score of 0 should be allocated (provided the studies is assessing resting-state EEG, and no rational for shorter epochs is provided in a task-related study). A score of 0.5 should be allocated if epoch are 4-6 seconds, and if epochs are greater than 6 seconds, a score of 1 should be allocated.

For Item 3, if epoch number is not mentioned, or the number of sample epochs assessed is less than 50, a score of 0 should be allocated. If 50-100 sample epochs are assessed, then a score of 0.5 should be allocated. If more than 100 sample epochs are used, then a score of 1 should be allocated. It should also be noted whether or not the same number of epochs was used in the different conditions being compared; this may be especially important for studies with fewer epochs. Furthermore, as discussed previously, the number of epochs in combination to the length of the window can impact the outcome of results. Optimal designs require more epochs for shorter windows when higher frequencies are assessed, and potentially less epochs with longer windows when lower frequencies are assessed. However, the authors still suggest using the criteria above for Item 3, irrespective of epoch length. With the 'ideal' epoch length being recommended as 6 second windows, achieving a score on both epoch length and number (>100) would generally require a recording of ~12 minutes (after artifact removal), which in the context of an EEG study does not seem unreasonable.

For Item 4, if no account of artifact rejection is provided, then a score of 0 should be allocated. If the study only mentions that artifacts including eye blinks, muscle movements, and electrical activity were removed but no methodology is described or the methodology has not been confirmed by the literature to remove more than 70% of artifact noise and leave more than 60% of the signal, then a score of 0.5 should be allocated. Lastly, if the study mentions that all 3 artifact types were removed (either using ICA or similar mathematical artifact reduction methods) or the technique used to clean that data is named, and the literature shows this technique removes more than 70% of artifact and leaves more than 60% of the signal (i.e., tools presently meeting this threshold include manual rejection, HAPPE, and MARA; Gabard-Durnam et al., 2018) then a score of 1 should be allocated.

Where in doubt, if a study does not provide enough information to discriminate between two scores on the checklist, it should be given the lower quality rating for that checklist item, given that it cannot be confirmed to meet the higher quality rating. Lastly, while there is suggestion from a single study, that using ICA to subtract artifactual components adversely affects the connectivity analysis, this finding needs to be confirmed with further research. If the study has used ICA to subtract noise, then this should be noted as a potential issue, but the study should still be marked according to the criteria above. The work to determine how significant the effect of component deletion or filtering should continue where possible, and the checklist proposed in this article should be updated to account for those findings.

For Item 5, if an article does not mention controlling for volume conduction (VC) and the EEG connectivity measure itself does not control for VC then a score of 0 should be allocated. If the chosen connectivity analysis controls for VC on its own (i.e., using zero-phase delay measures, imaginary Coh, partial directed Coh, or weighted PLI to name a few), or if the study employs extra controls such as imposing phase-lags and surface Laplacian, individually, then a score of 0.5 is warranted. However, if *both* the connectivity measure chosen controls VC and further steps (like surface Laplacian) have been applied, then a score of 1 should be provided to the study on this item.

For Item 6a, if multiple comparisons are made but are not controlled for then a score of 0 should be provided. If the study reports or uses a more stringent p-value such as 0.01, a Bonferroni correction or FDR where mass-univariate testing takes place, then a score of 0.5 should be provided. If multiple comparisons are controlled for by limiting the analysis to whole-brain or specific pairs, thus limiting the number of comparisons, and the Bonferroni or FDR methods are used, then a score of 1 should be allocated; the analysis and specific pairs are named prior to statistical testing.. Further, where mass-univariate testing is employed and non-parametric permutation statistics (i.e., cluster-base statistics or NBS) are used, do not score 6a, instead use 6b.

Item 6b should be used in the place of Item 6a where network metrics and analyses are used to assess the topography of connections. In which case, if thresholds for network analysis are set arbitrarily, a score of 0.5 should be provided. Whereas, if objective, model-driven thresholds are met or MST is used where a threshold is not necessary and/or weighting is employed, a score of 1 should be provided.

Lastly, for Item 7 there are considerable discrepancies in sample size effects, selection, and analysis in the literature, and no clear criteria or recommendations can be made about a specific number. The authors encourage researchers to increase attention to the reporting and performing of a priori sample size calculations and thus for the present checklist. If the research paper reports the inclusion of a sample size based on statistical considerations, including sample sizes considered and calculated from extant studies or pilot data, sample sizes calculated from software's such as G*Power or Stata, then a score of 1 should be provided to the study. If the research paper makes no mention of how they came to their desired sample size, and it does not appear that any prior sample size considerations were made, then a score of 0 should be awarded.